\documentclass[reprint,superscriptaddress,amsmath,amssymb,aps]{revtex4-2}
\usepackage{subcaption}
\usepackage{natbib}
\usepackage[colorlinks,linkcolor = blue,citecolor=blue,urlcolor=blue,bookmarks=false,hypertexnames=true]{hyperref}
\usepackage{float}
\usepackage{comment}
\usepackage{mathtools}
\usepackage{graphicx}% Include figure files
\usepackage[export]{adjustbox}
\usepackage{dcolumn}% Align table columns on decimal point
\usepackage{bm}% bold math
\usepackage{sidecap}

\usepackage{caption}
\usepackage{ragged2e} % gives advanced justification options

\usepackage{chemformula}
\usepackage[percent]{overpic}
\usepackage{gensymb}

\begin{document}

\title{Magnomechanical Coupling in Suspended 2D van der Waals Ferromagnets}

\author{Ritesh Das}
\email{R.D.Das@tudelft.nl}%Lines break automatically or can be forced with \\
\affiliation{%
Kavli Institute of Nanoscience, Delft University of Technology, Lorentzweg 1, 2628 CJ, Delft, The
Netherlands
}%

\author{Alvaro Bermejillo-Seco}
\affiliation{%
Kavli Institute of Nanoscience, Delft University of Technology, Lorentzweg 1, 2628 CJ, Delft, The
Netherlands
}%

\author{Herre S. J. van der Zant}
\affiliation{%
Kavli Institute of Nanoscience, Delft University of Technology, Lorentzweg 1, 2628 CJ, Delft, The
Netherlands
}%

\author{Peter G. Steeneken}
\affiliation{%
Kavli Institute of Nanoscience, Delft University of Technology, Lorentzweg 1, 2628 CJ, Delft, The
Netherlands
}%
\affiliation{%
Department of Precision and Microsystems Engineering, Delft University of Technology, Mekelweg 2, 2628 CD, Delft, The Netherlands
}%

\author{Yaroslav M. Blanter }
\email{Y.M.Blanter@tudelft.nl}%Lines break automatically or can be forced with \\
\affiliation{%
Kavli Institute of Nanoscience, Delft University of Technology, Lorentzweg 1, 2628 CJ, Delft, The
Netherlands
}%

\date{\today}

\begin{abstract}
Magnomechanical systems provide a promising route for exploring coherent hybrid magnon–phonon interactions and hybrid information processing, but their realization has so far been limited by weak magnon–phonon coupling in conventional bulk platforms. We show that a suspended membrane of a two-dimensional van der Waals ferromagnet with in-plane magnetization and out-of-plane mechanical oscillations exhibits large magnomechanical coupling dominated by magnetoelastic interactions. The parametric single magnon–phonon coupling rate scales linearly with pre-strain and can reach hundreds of Hertz to low kiloHertz in suspended membranes of van der Waals ferromagnets such as \ch{CrGeTe3} under experimentally realistic conditions. This rate exceeds typical values reported for YIG spheres by more than three orders of magnitude. Our results demonstrate that suspended membranes of van der Waals magnets provide a robust and highly tunable platform for magnomechanics.
\end{abstract}

\maketitle
\section{Introduction}
The rapidly developing field of two-dimensional (2D) van der Waals (vdW) materials has unlocked unprecedented opportunities for exploring novel quantum phenomena at the atomic scale \cite{vd0, vd1, vd2}. Among these, 2D magnets \cite{vdw1,vdw2,vdw3,vdw4, roadmap} have emerged as a compelling platform for investigating fundamental aspects of magnetism, spin dynamics, and their coupling to lattice vibrations \cite{2d1,2d2,jaimecgt}. The ability to achieve stable magnetic ordering in monolayer and multilayer systems \cite{cgt,cri} has spurred intense interest in the interplay between magnetic, electronic, and mechanical degrees of freedom in reduced dimensions, particularly in widely studied systems such as \ch{CrI3} \cite{firstcri, cri} and \ch{CrGeTe3} \cite{cgt}.

For example, magnons and phonons can interact coherently through strain- or motion-induced modulation of magnetic order, an effect broadly studied under the field of magnomechanics \cite{clinton, mewaves, heikkila, cavitymagno, magnomech, magnomech2}. The interaction is closely related to optomechanics \cite{aspelmeyer2014cavity, quantumem,quantumtel}, where mechanical motion modulates an optical cavity to couple photons and phonons. In magnomechanics, the role of the optical cavity is replaced by a magnetic material which acts as a magnon cavity \cite{heikkila}. This coupling provides a route to controlling spin excitations with mechanical motion, enabling applications ranging from hybrid quantum systems \cite{zuo2024cavity} and mechanically mediated magnonics \cite{clinton} to precision sensing, transduction \cite{hatanaka2022chip, optimaltrans,Bondarenko2026Magnetoelastic}, and quantum storage devices \cite{magnomechquantum}.

To date, the majority of experimental work has been carried out in three-dimensional platforms, most notably in YIG (yttrium-iron-garnet) resonators, which are commonly fabricated as spheres that are hundreds of microns in diameter \cite{clinton, cavitymagno}. These bulk geometries support low magnon damping but inherently limit zero-point motion and generate only weak strain for a given displacement, leading to single magnon–phonon coupling rates of tens of milliHertz \cite{clinton, cavitymagno}. The spherical shape and macroscopic size also hinder integration with chip-scale devices. A first step towards more compact platforms was taken by Kansanen \textit{et al.} \cite{heikkila}, who theoretically examined a suspended \ch{CoFeB} beam, approximately 50 $\mu$m long and 10 $\mu$m wide. Despite its reduced thickness and more planar geometry, the predicted single–magnon–phonon coupling strengths remain in the 10–100 mHz range, similar to those of YIG spheres, and are still constrained by the overall scale of the mechanical resonator and the mechanism of coupling. 

For quantum applications, the relevant figure of merit is the cooperativity, which must exceed unity in order to reach the regime of coherent magnon–phonon dynamics. In both optomechanical and magnomechanical systems, achieving this condition generally relies on strong external driving to enhance the interaction through linearization. In the absence of driving, the single magnon-phonon cooperativity is very small (of the order of $\sim 10^{-12}$) for the examples mentioned above. Realizing larger intrinsic single–magnon–phonon coupling strengths significantly lowers the required drive power, easing experimental requirements and bringing the quantum regime within closer reach.

Membranes made of thin vdW magnets are well positioned to overcome the geometric limitations of micrometer-sized spheres and beams. Their nanometer-scale thickness and micron-scale lateral dimensions \cite{makars} naturally increase zero-point motion and strain, suggesting the possibility of significantly stronger magnon–phonon coupling than in suspended beams of bulk magnets \cite{heikkila}. These membranes can be readily integrated into chip-scale architectures and offer a highly tunable platform for both magnetic and mechanical control \cite{cgtms,onchip}. Additionally, suspending vdW materials often enhances their intrinsic properties; for instance, suspended graphene shows markedly increased carrier mobility \cite{suspendedgraphene}, hinting at similar benefits in other materials. The suspended-membrane approach has likewise been applied to 2D magnets \cite{mak1, makars, bermejillo2025thermoelastic,maurits}, enabling precise determination of their magnetic transition temperatures \cite{makars}. Yet, despite these advantages of vdW membranes as a platform for magnomechanics, magnon–phonon coupling in suspended vdW magnetic membranes has, to our knowledge, not yet been explored.

In this work, we propose a suspended doubly-clamped two-dimensional membrane composed of a thin (few-layer) ferromagnet as a versatile platform for magnomechanics. A similar geometry has been used to study and quantify magnetostriction in thin films \cite{hans1,hans2}. An external magnetic field is applied to align the spins within the plane of the membrane, thereby maximizing their coupling to the membrane’s mechanical modes. As a representative model system, we focus on \ch{CrGeTe3}, a hexagonal van der Waals ferromagnet with an out-of-plane easy axis \cite{cgt}. While our analysis is presented in the context of this material, the results are general and readily extend to other two-dimensional ferromagnets. We select \ch{CrGeTe3} in particular due to its experimental accessibility and large magnetoelastic coupling \cite{cgtmagnomech1, cgtmagnomech2}.

We model a suspended magnetic membrane oscillating out of plane and find that finite initial deformation is required for nonzero magnomechanical coupling, consistent with the findings of Kansanen \textit{et al.}~\cite{heikkila}. In buckled beams the coupling originates from second-order strain associated with the static curvature of the structure \cite{heikkila}. By contrast, in suspended two-dimensional membranes, out-of-plane deflection induces first-order modulation of the in-plane strain. We show that the single magnon-phonon coupling in our platform can reach values in the order of few hundred Hz to kHz, which is more than three orders of magnitude higher than those observed in YIG spheres and \ch{CoFeB} beams.

\begin{figure}[]
 \begin{center}
    \begin{subfigure}{0.4\textwidth}
    \begin{overpic}[width=\linewidth]{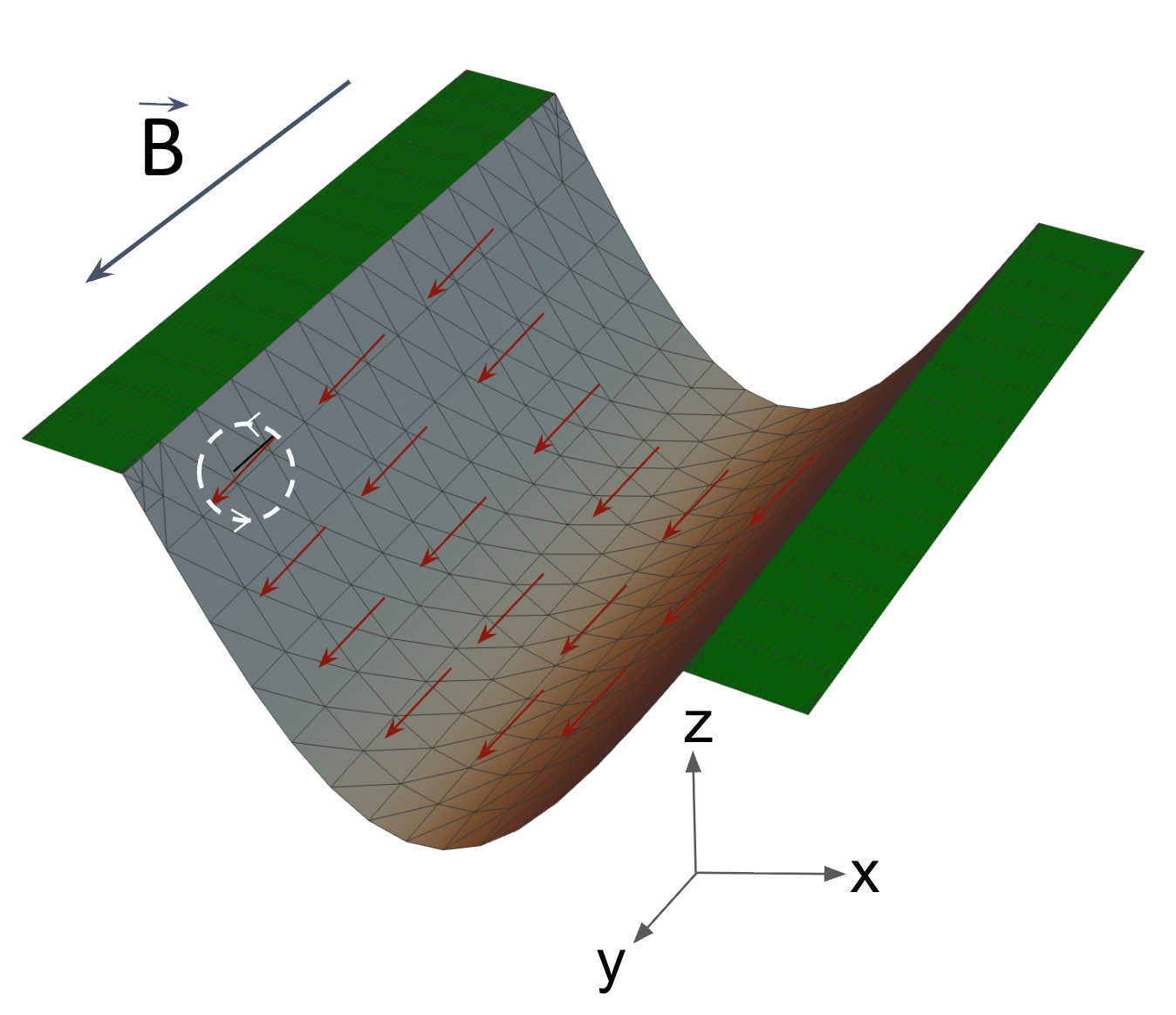}
        \put(2,80){\large (a)}
    \end{overpic}
  \end{subfigure}\\
  \begin{subfigure}{0.4\textwidth}
    \begin{overpic}[width=\linewidth]{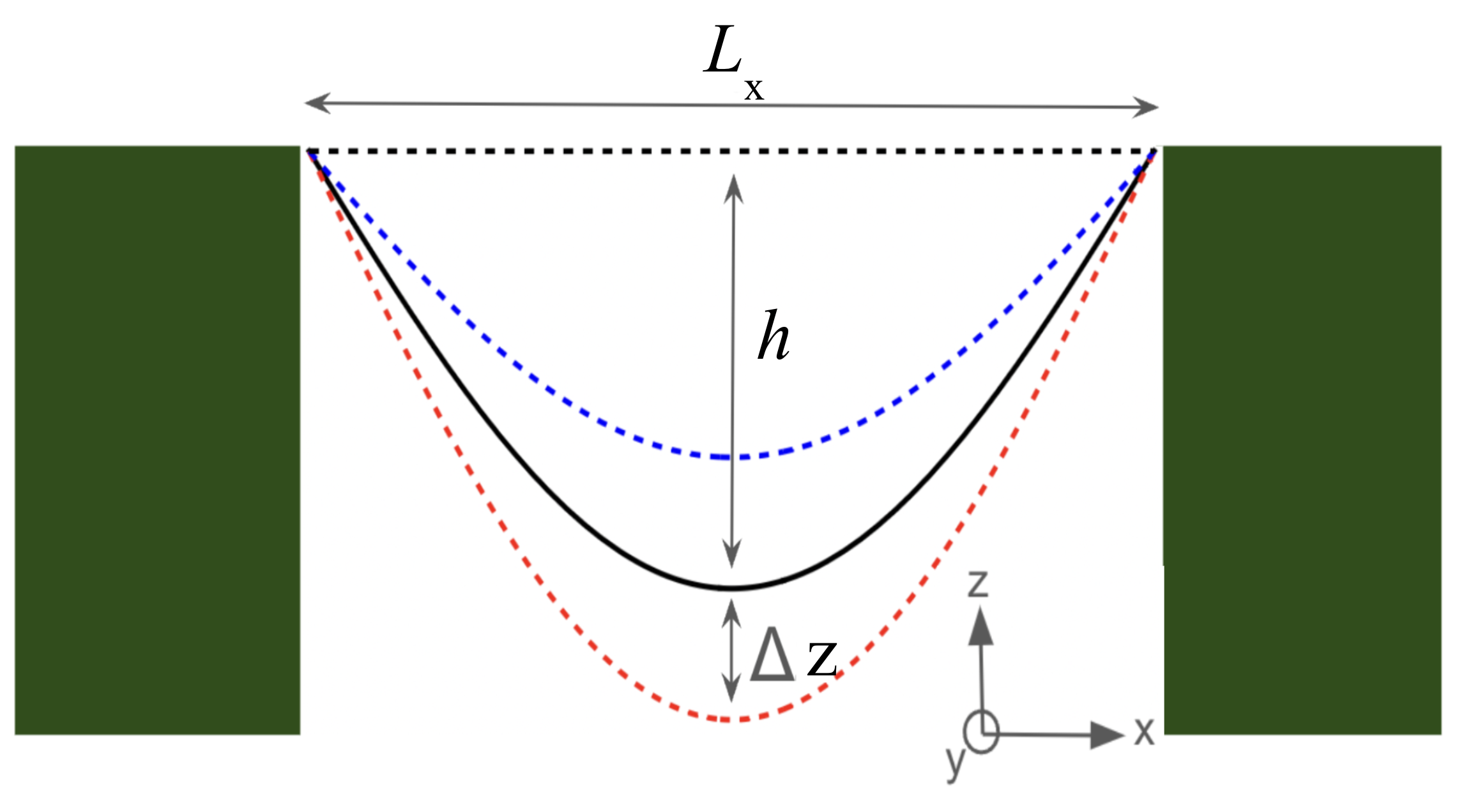}
        \put(2,50){\large (b)}
    \end{overpic}
  \end{subfigure}
  \end{center}
 \caption{The magnomechanical system. (a) 3D view of the suspended ferromagnet. The magnons are aligned in-plane via an external magnetic field along $\text{y}$. (b) 2D cross-section of the magnomechanical system. The magnet is pre-strained by pulling it down vertically by a distance $h$ and oscillates around this position with amplitude $\Delta \text{z}$. The mechanical displacements are exaggerated for clarity.}
\end{figure}

The paper is organized as follows: Section II describes the design of the model and our motivation for adopting the specific model. Section III discusses the Hamiltonian of the system and the theory of magnomechanical coupling. We derive an analytical expression for the single magnon-phonon coupling. In Section IV, we display and analyze the results. We finish with the conclusions in Section V.
\section{Model}
Our proposed magnomechanical platform consists of a 2D membrane of a ferromagnetic material suspended over a cavity. The magnetic material behaves as a suspended doubly-clamped thin membrane (shown in Figure 1) that oscillates normal to the plane. Throughout this work, the term phonons denotes the discrete out-of-plane mechanical modes of the suspended membrane. We focus on the fundamental mode of the membrane, assuming a uniform out-of-plane oscillation. The lateral in-plane oscillations along the width are considered to be energetically unfavorable. The extreme positions of the membrane oscillating in the fundamental mode are shown in Figure 1 (b).

We consider a membrane of length $L_\text{x}$, width $L_\text{y}$, and thickness $L_\text{z} \ll L_\text{x}$ (see Figure 1). Suspended membranes with micrometer-scale lateral dimensions and nanometer-scale thicknesses typically exhibit fundamental mechanical mode frequencies $\omega_b$ in the range of $10$--$100~\mathrm{MHz}$, with quality factors on the order of $10^4$ \cite{makars}. These relatively high resonance frequencies arise from the small membrane dimensions, as the mechanical frequency scales inversely with the system size. To obtain a nonzero linear magnon-phonon coupling, it is necessary to break the symmetry of the time-dependent strain \cite{heikkila}. In the present geometry, this is achieved by introducing a static out-of-plane deformation $h$ of the membrane (Figure 1), for example via electrostatic actuation using a gate voltage \cite{wang2010deflection}. The membrane is initially displaced by a height $h$ from the center, thereby generating a static in-plane tensile strain, and oscillated with an out-of-plane amplitude $\Delta \text{z} \ll h$. 

This approach differs fundamentally from the pre-strain induced by beam buckling considered by Kansanen \textit{et al.}~\cite{heikkila}. In the buckled-beam geometry, the coupling relies on a second-order strain contribution associated with the buckled shape. In contrast, for a suspended two-dimensional membrane, a static out-of-plane deformation directly produces an in-plane tensile strain. Small oscillations about this deformed configuration generate a first-order, time-dependent modulation of the in-plane strain due to the extension and compression along the $x-$axis. This dynamic strain couples efficiently to the magnetoelastic energy. Because this mechanism does not depend on strain gradients across a finite thickness, it remains effective in the thin limit and is strongly enhanced for small, lightweight membranes, leading to substantially larger single-magnon-phonon coupling strengths.

\begin{figure}[]
  \centering
      \includegraphics[width=\linewidth]{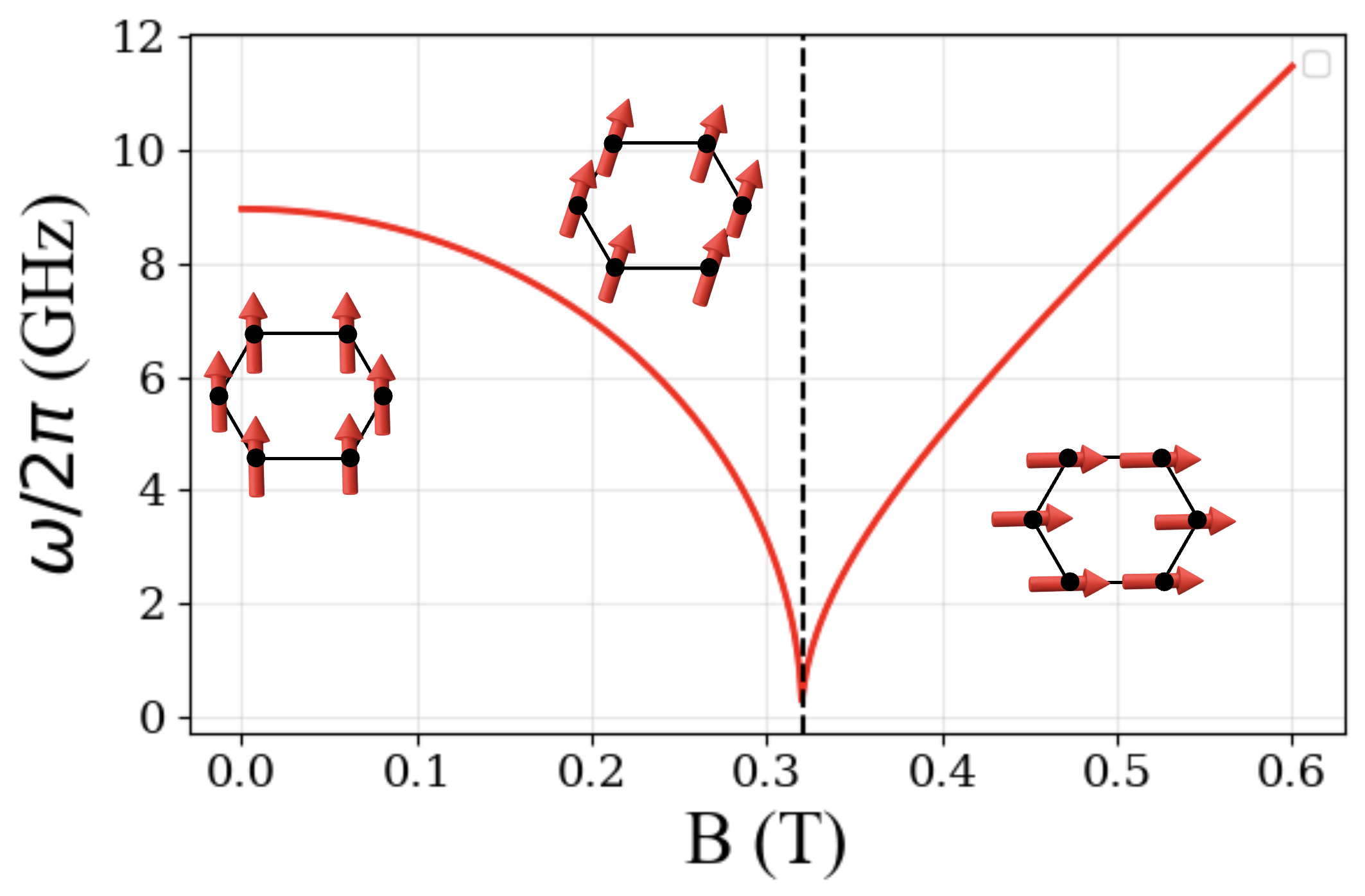}
  \caption{Ferromagnetic resonance frequency of \ch{CrGeTe3} vs. an applied in-plane external field. The insets show the ground state of \ch{CrGeTe3} at the corresponding field. The external field cants the spins towards the plane. The dashed line represents the saturation field at which the spins lie completely in-plane \cite{cgtfmrtheory, cgtms}.}
\end{figure}

While our results are general, we present numerical values for \ch{CrGeTe3}, a hexagonal ferromagnet with an observed Gilbert damping parameter $\alpha_{M} \sim 4\text{--}10 \times 10^{-4}$ in thick flakes \cite{gildamp}. To obtain magnomechanical coupling, it is essential to have an overlap between the dynamic magnetization and the strain field. Such overlap can arise from spin precession about either the $\text{z}$- or $\text{y}$-axis. Moreover, precession about the $\text{y}$-axis can, in principle, provide an additional contribution from demagnetizing fields, although, as we show later, this contribution is very small. For completeness, we present the configuration with magnetization aligned along $\text{y}$, noting that alignment along $\text{z}$ yields nearly the same coupling strength.

Figure~2 shows the ferromagnetic resonance frequency of the lowest magnon mode in \ch{CrGeTe3} as a function of an in-plane magnetic field applied along the $\text{y}$-axis (perpendicular to the length of the membrane)  \cite{cgtms}. The insets illustrate the evolution of the magnetic structure with increasing field. We apply a magnetic field larger than $\sim$300~mT to fully saturate the spins in-plane. When driven, the spins precess about the $\text{y}$-axis, generating magnetization components along the $x$- and $\text{z}$-directions (see dashed circle in Figure~1 (a)). In this configuration, the magnon oscillations overlap with the tensile strain oscillations along $x$.

We focus on the Kittel mode, the lowest-frequency magnetic resonance mode. Close to the saturation field, magnons are unstable, but they become stable at slightly higher fields, exhibiting GHz-scale frequencies that are significantly higher than those of the mechanical modes.

\section{Theory}
\subsection{Hamiltonian}

To describe the magnon-phonon coupling, we need to write the free energy density in terms of quantized operators of the magnetization and the mechanical mode. The magnetization can be written in terms of bosonic magnon operators via the linearized Holstein-Primakoff transformation, while the displacement field is expanded and quantized using phonon operators (see Appendix A). We start by writing the free energy density for the system. The magnetoelastic free energy density is given by \cite{heikkila,clinton}

\begin{equation}
\begin{aligned}
\mathcal{F}_{\text{me}} &= \frac{b_{1}}{M_s^2} \left( M_\text{x}^2 \epsilon_{\text{xx}} + M_\text{y}^2 \epsilon_{\text{yy}} + M_\text{z}^2 \epsilon_{\text{zz}} \right) \\
&\quad + \frac{2b_{2}}{M_s^2} \left( M_\text{x} M_\text{y} \epsilon_{\text{xy}} + M_\text{x} M_\text{z} \epsilon_{\text{xz}} + M_\text{y} M_\text{z} \epsilon_{\text{yz}} \right).
\end{aligned}
\end{equation}
Here, $M_{s}$ is the saturation magnetization, $b_{1}, b_{2}$ are the magnetoelastic constants, $M_{i}$ is the magnetization along $i$, where $i \in (\text{x,y,z})$, and $\epsilon_{ij}$ is the strain tensor given by

\begin{equation}
\epsilon_{ij} = \text{x}_{\text{zpf}} \left( \hat{b} + \hat{b}^{\dagger} \right) \frac{1}{2} \left( \frac{\partial u_i}{\partial x_j} + \frac{\partial u_j}{\partial x_i} \right),
\end{equation}
where $\hat{b} (\hat{b}^{\dagger})$ is the phonon annihilation (creation) operator, $x_{i}$ is the displacement coordinate along $i$, $\text{x}_{\mathrm{zpf}}$ is the zero point fluctuations, and $u_{i}$ is the component of the unitless normalized displacement profile \cite{clinton} along direction $i$ such that $\text{r}_{i} = \text{x}_{\text{zpf}} ( \hat{b} + \hat{b}^{\dagger} )u_{i}$, where $\text{r}_{i}$ is the displacement field along $i$.

The demagnetizing free energy density terms (that do not cancel out upon integrating over the volume) in the limit $L_{\text{z}} \ll L_{\text{y}} < L_{\text{x}}$ are given by \cite{heikkila}

\begin{equation}
   \mathcal{F}_{\text{dm}} = \mu_{0}\frac{M_\text{z}^2}{2} +\mu_{0}(M_\text{x}^2 - M_\text{z}^2)\epsilon_{\text{xx}}.
\end{equation}
Since the magnetization is saturated along $\text{y}$, the time-dependent components of the total magnetization are along $\text{x}-$ and $\text{z}-$directions. They can be linearized and written in terms of the magnon creation (annihilation) operators $\hat{m}(\hat{m}^{\dagger})$ as
\begin{equation}
\begin{split}
M_\text{x} &= \sqrt{\frac{\hbar \gamma M_s}{2 V_\text{m}}} \left( \hat{m} + \hat{m}^\dagger \right), \\
M_\text{z} &= i \sqrt{\frac{\hbar \gamma M_s}{2 V_\text{m}}} \left( \hat{m} - \hat{m}^\dagger \right),
\end{split}
\end{equation}
where $\gamma$ is the gyromagnetic ratio and $V_{\text{m}}$ is the total volume of the membrane. As discussed in Sec. II, the membrane has a static deformation of height $h$ and is clamped at both ends, $(-L_{\text{x}}/2, 0)$ and $(L_{\text{x}}/2,0)$. In the limit $\Delta \text{z}, L_{\text{z}} \ll h, L_{\text{x}}$, the dominant strain arises from first-order extension and compression along the length of the membrane, represented by $\epsilon_{\text{xx}}$. Other components such as $\epsilon_{\text{xz}}$ or $\epsilon_{\text{zz}}$, appear only at second order in $\Delta \text{z} / L_\text{x}$ or $L_\text{z} / L_\text{x}$, and are therefore neglected. The time-dependent component of $\epsilon_{\text{xx}}$ is given by

\begin{equation}
\epsilon_{\text{xx}}(t) = \frac{\pi^{2} h}{L_\text{x}^2} \sqrt{\dfrac{\hbar}{2m_{\text{eff}} \omega_{b}}}\sin^{2} \left[ \dfrac{\pi \text{x}}{L_{\text{x}}} \right] (\hat{b}^\dagger + \hat{b}) = \epsilon^{\text{max}}_{\text{xx}} (\hat{b}^{\dagger} + \hat{b}),
\end{equation}
where $m_{\text{eff}} = \int_{V}\rho |u|^{2}dV = \rho V_{\text{m}}/2$ is the effective mass, and $\epsilon^{\text{max}}_{\text{xx}}$ is the amplitude of the strain profile $\epsilon_{\text{xx}}$. For uniform magnon modes (such as the Kittel mode), the exact variation of the strain with position is not important and can be replaced with the average global strain \cite{clinton}. 

We can write the total magnomechanical Hamiltonian by substituting Eqs. (4) and (5) in Eqs. (1) and (3) and integrating over volume. This results in
\begin{equation}
    H_{mb} = g_{\text{0}} \hat{m}^{\dagger}\hat{m} (\hat{b}+ \hat{b}^{\dagger}),
\end{equation}
where $g_{\text{0}} = g_{\text{me}} + g_{\text{dm}}$ denotes the total single-magnon–phonon coupling rate; $g_{\text{me}}$ arises from the magnetoelastic interaction and $g_{\text{dm}}$ from demagnetizing-field effects. In addition, shear components in Eq. (1) can generate a cross term linear in the magnon operators, of the form $(\hat{m} + \hat{m}^{\dagger})(\hat{b} + \hat{b}^{\dagger})$. As discussed in Sec. II, for the systems considered here, the magnon frequency $\omega_{m}$ is far from the phonon frequency $\omega_{b}$. Consequently, such linear terms are nonresonant and are neglected in Eq. (6).

By contrast, the parametric interaction term $\hat{m}^{\dagger} \hat{m} (\hat{b} + \hat{b}^{\dagger})$ is non-linear in the magnon operators and becomes resonant under appropriate driving conditions. This parametric coupling may be linearized, yielding an effective coupling strength $g_{\text{eff}} = g_{\text{0}} \sqrt{\langle \hat{m}^{\dagger} \hat{m} \rangle}$ \cite{clinton}. In this work, we do not further consider the driven regime or the resulting effective coupling.

\begin{figure*}
 \begin{center}
    \begin{subfigure}{0.475\textwidth}
    \begin{overpic}[width=\linewidth]{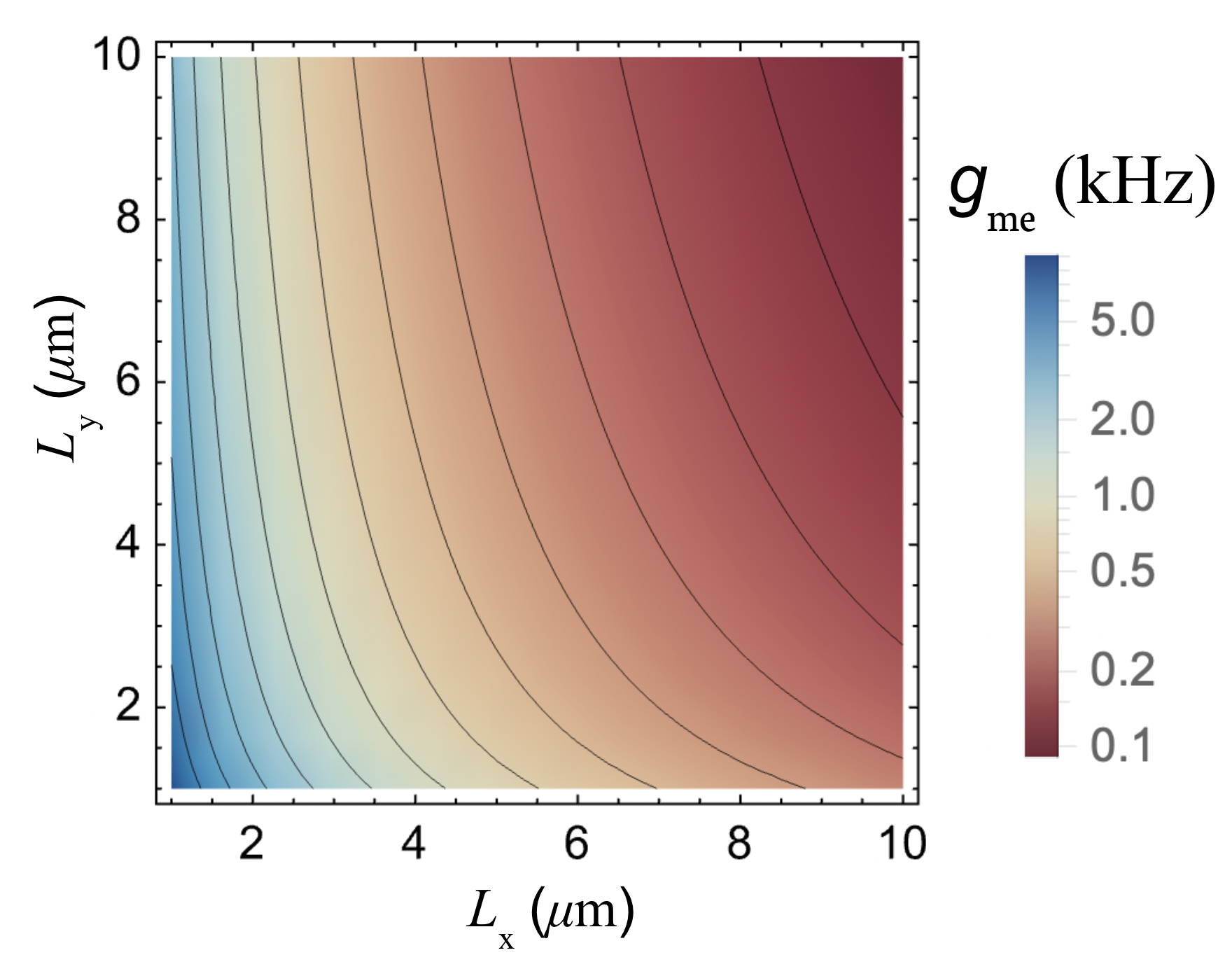}
        \put(65,67){\large (a)}
    \end{overpic}
  \end{subfigure}
  \begin{subfigure}{0.47\textwidth}
    \begin{overpic}[width=\linewidth]{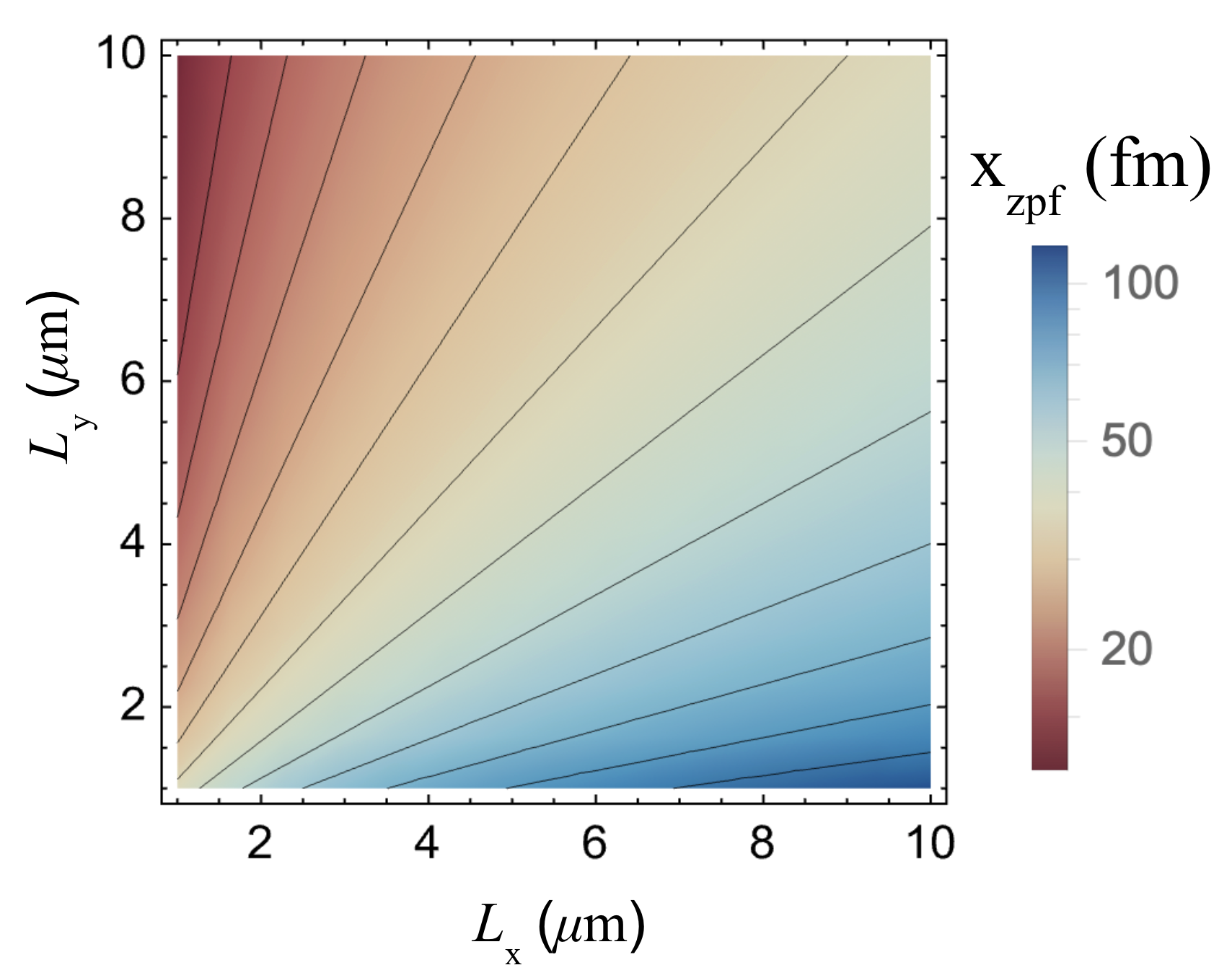}
        \put(67,69){\large (b)}
    \end{overpic}
  \end{subfigure}
  \end{center}
 \caption{(a) The magnetoelastic single magnon-phonon coupling rate $g_{\text{me}}$ and (b) zero point fluctuation $\text{x}_{\mathrm{zpf}}$ vs. membrane length $L_{\text{x}}$ and membrane width $L_{\text{y}}$ for initial deformation $h = 100$ nm and thickness $L_{\text{z}}=20$ nm. The contour lines are for visual guidance only.}
\end{figure*}

\begin{figure}
 \begin{center}
    \includegraphics[width=\linewidth]{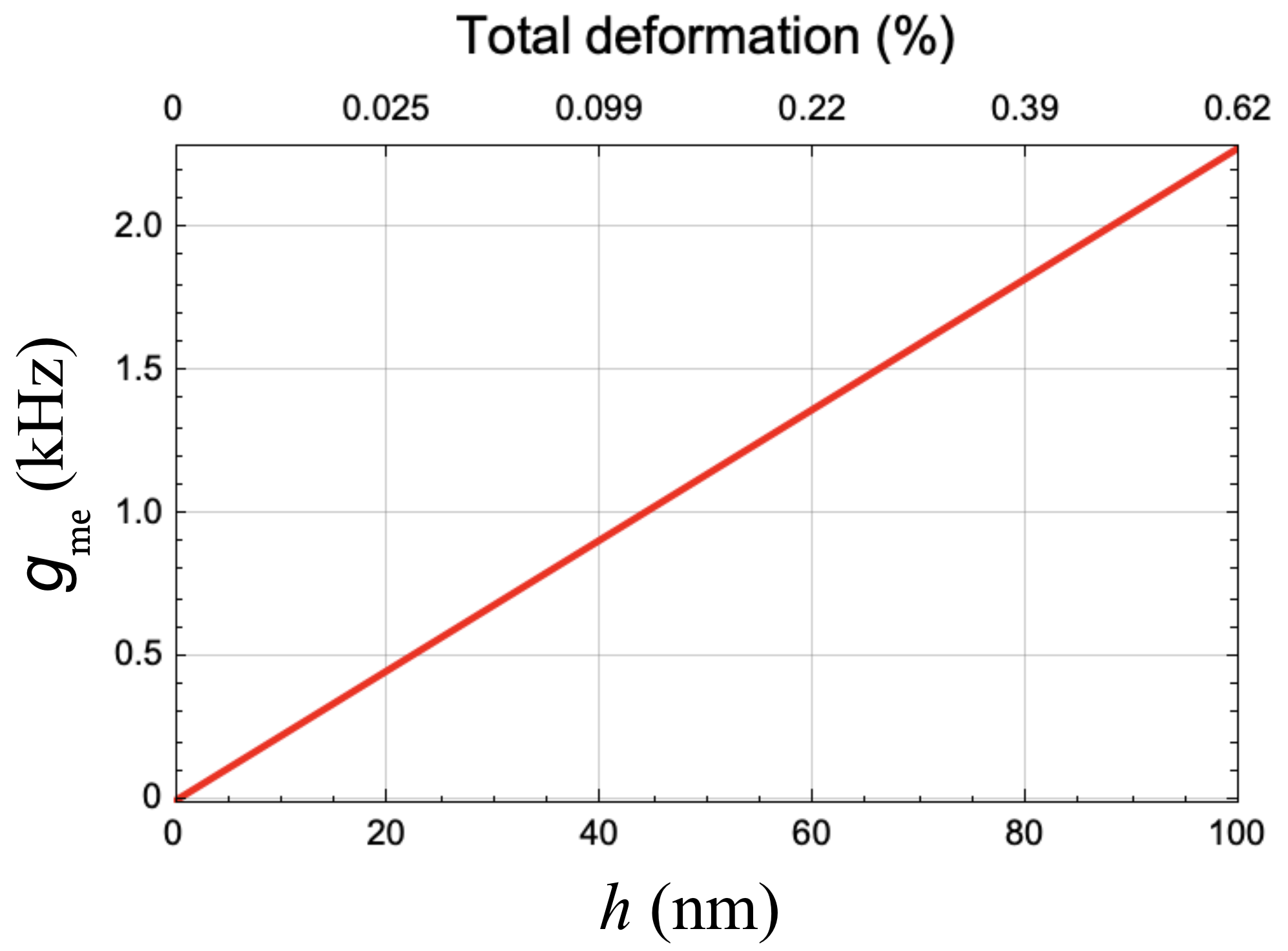}
    \caption{The magnetoelastic single magnon-phonon coupling rate $g_{\text{me}}$ vs. initial displacement and total deformation for $L_{\text{x}} = L_{\text{y}} = 2$ $\mu$m and $L_{\text{z}}=20$ nm.} 
  \end{center}
 \end{figure}

\subsection{Magnon-Phonon Coupling}
We begin by calculating the magnetoelastic coupling $g_{\text{me}}$. Integrating the free energy density given in Eq. (1) and isolating only the non-linear terms as in Eq. (6), the total magnetoelastic free energy is found as
    \begin{equation}
   E_{\text{M}} = \int_{V} \dfrac{b_{1}}{M^{2}_{s}}(M^{2}_{\text{x}} \epsilon_{\text{xx}}) dV.
\end{equation}
Substituting $M_{\text{x}}$ from Eq. (4) and $\epsilon_{\text{xx}}$ from Eq. (5), we get
\begin{equation}
    E_{\text{M}} = \dfrac{\hbar  b_{1} \gamma}{ M_{s} V_{\text{m}}} \int_{V} \epsilon^{\text{max}}_{\text{xx}} \, d\text{x} \, d\text{y} \, d\text{z}.
\end{equation}
Comparing Eq. (8) with Eq. (6) and using $M_{s} = \hbar \gamma n_s$ \cite{stancilprabhakar, clinton}, we find the single magnon-phonon coupling rate due to magnetoelastic interaction as
\begin{equation}
    g_{\text{me}} = \dfrac{ b_{1}}{\hbar V_{\text{m}} n_{s}}\int^{L_{\text{x}}/2}_{-L_{\text{x}}/2} \int^{L_{\text{y}}/2}_{-L_{\text{y}}/2} \int^{L_{\text{z}}/2}_{-L_{\text{z}}/2} \epsilon^{\text{max}}_{\text{xx}} d\text{x} d\text{y} d\text{z}.
\end{equation}
Calculating, we get
\begin{equation}
    g_{\text{me}} = \dfrac{ b_{1}}{ \hbar n_{s}}\frac{\pi^2}{2} \frac{h}{L_\text{x}^2}\sqrt{\dfrac{\hbar}{2m_{\text{eff}} \omega_{b}}}  ,
\end{equation}
where $n_{s}$ is the spin density. Thus, we find that $g_{\text{me}}$ depends linearly on the initial deformation $h$.

Similarly, by integrating the demagnetizing free energy density in Eq. (3), we can calculate 

\begin{equation}
    g_{\text{dm}} = \dfrac{\mu_{0} M_{s}^{2}}{ \hbar n_{s}}\frac{\pi^2}{2} \frac{h}{L_\text{x}^2}\sqrt{\dfrac{\hbar}{2m_{\text{eff}} \omega_{b}}} = \dfrac{\mu_{0}M_{s}^{2}}{b_{1}}g_{\text{me}}. 
\end{equation}
The factor $\alpha = \mu_{0}M_{s}^2/b_{1}$ is a measure of the strength of demagnetizing fields compared to the magnetoelastic interaction.
\section{Analysis}    
The behavior of the single magnon-phonon coupling $g_{\text{0}}$ is analyzed as a function of the geometric parameters of the membrane and the initial deformation. Eqs. (10) and (11) show that the presence of an initial deformation is essential for achieving a non-zero magnomechanical coupling \cite{heikkila}. Further, the coupling increases with decreasing system size and depends on $\alpha$. For $\alpha > 1$, the demagnetizing fields dominate the magnon-phonon coupling rate, while for $\alpha < 1$, the magnetoelastic interaction dominates. The parameter $\alpha$ is an intrinsic material property and exceeds unity in ferromagnets such as \ch{CoFeB} with large saturation magnetization \cite{heikkila}, while it is smaller than unity in materials with relatively low saturation magnetization and strong magnetoelastic coupling, such as YIG \cite{cavitymagno}. For vdW magnets, there is no direct measurement of the magnetoelastic constant yet, to the best of our knowledge. However, large magnetoelastic coupling has been observed in \ch{CrGeTe3} \cite{cgtmagnomech1, cgtmagnomech2}. This is further supported by previous DFT calculations of the strain-dependent magnetic  exchange constants in CrGeTe$_3$~\cite{jaimecgt}. From the reported variation of the exchange constants with applied strain in Ref. \cite{jaimecgt}, we can extract an order of-magnitude estimate $b_{1} \sim 10^5 - 10^6$ Pa. For numerical analysis, we take $b_{1} = 3 \times 10^5$ Pa, the same value as reported in YIG \cite{clinton}, spin density $n_{s} \approx 6.7 \times 10^{27}$m$^{-3}$, and density $\rho = 5.68$ g.cm$^{-3}$. This results in $\alpha \approx 0.05$, and $g_{\text{me}} \approx 20g_{\text{dm}}$. The spin density is roughly of the same order in other vdW magnets. Thus, the dominant mechanism of coupling in vdW magnets is the magnetoelastic interaction. In subsequent numerical analysis, we focus on $g_{\text{me}}$. Finally, we note that the magnomechanical coupling is tunable through the membrane strain. Since the coupling strength scales with the pre-strain, it can be enhanced by increasing the gate-induced deformation or by engineering the built-in strain during fabrication.

\begin{table*}[]
\centering
\begin{tabular}{lcccccccc}
\hline
Platform & $g_{\text{dm}}$ & $g_{\text{me}}$ & $\kappa_{\text{m}}$ & $\kappa_b$ & Size  & $\alpha$ & $C_{\text{sm}}$ \\
\hline
YIG sphere \cite{clinton} 
& $\sim$ mHz 
& 30 mHz 
& $6.28$ MHz 
& 1.8 kHz 
& $0.01\,$mm$^{3}$ sphere  
& $ 0.065$ 
& $3.18 \times 10^{-13}$ \\

CoFeB beam \cite{heikkila} 
& 150 mHz 
& $\sim$ mHz 
& $\sim100$ MHz 
& $\sim$ kHz 
& $50\times 10 \times 0.1\,\mu\mathrm{m}^{3}$  
& $\sim 10^{2}$ 
& $ \sim1 \times10^{-12}$ \\

CrGeTe$_3$ membrane (This work) 
& $\sim$ Hz 
& $ 5$ kHz 
& $ \sim 0.6$ GHz \cite{cgtms} 
& $\sim$ kHz 
& $1\times 1 \times 0.020\,\mu\mathrm{m}^{3}$  
& $0.05$ 
& $ 1.6 \times 10^{-4}$ \\
\hline
\end{tabular}
\caption{Comparison of YIG spheres \cite{clinton}, CoFeB beams \cite{heikkila}, and CrGeTe$_3$ membranes (this work) in terms of magnon--phonon coupling, dissipation rates, and cooperativity. The values are either taken or extracted from the corresponding references. The sign "$\sim$" denotes an estimate of the order of the quantity wherever it is highly variable or not given.}
\end{table*}

In Figure 3 (a), we plot $g_{\text{me}}$ as a function of $L_{\text{x}}$ and $L_{\text{y}}$ for $h = 100$ nm and $L_{\text{z}} = 20$ nm. The mechanical resonance frequency is taken to be $\omega_{b} = 2\pi \times 1$ MHz for a 10 $\times$10 $\mu$m$^2$ membrane and is inversely proportional to the square of length. The single magnon-phonon coupling rate reaches $\sim 5$ kHz for membranes with $L_{\text{x}} = L_{\text{y}} \approx 1~\mu\text{m}$ and $L_{\text{z}} = 20~\text{nm}$. For experimentally feasible dimensions of $L_{\text{x}} = L_{\text{y}} \approx 2~\mu\text{m}$, the coupling remains close to the kHz scale at the same deformation $h = 100$ nm (resulting in a total deformation of 0.62 \%), which is already two orders of magnitude larger than typical single photon-phonon coupling rates in optomechanical systems, and more than four orders of magnitude larger than the single magnon-phonon coupling rate in YIG spheres and CoFeB beams, indicating that even if $b_{1}$ were an order of magnitude smaller, the coupling would remain several orders of magnitude higher. The reported enhancement is therefore robust against the uncertainty in $b_1$. The large coupling arises from the first-order strain factor, $h/L^{2}_\text{x}$ in Eq. 10, and the increase in zero-point fluctuation $\text{x}_{\mathrm{zpf}}$ due to the reduced dimensions. To understand the specific contributions, we compare the behavior of the zero-point fluctuations with those in other devices.

In Figure 3 (b), we plot the zero-point fluctuation $\text{x}_{\mathrm{zpf}}$ as a function of $L_{\text{x}}$ and $L_{\text{y}}$ for the suspended membrane. The calculated $\text{x}_{\mathrm{zpf}}$ in suspended membranes is an order of magnitude larger than in \ch{CoFeB} beams~\cite{heikkila} and two orders of magnitude larger than in YIG spheres~\cite{cavitymagno}. Comparing Figure~3(b) with Figure~3(a), we find that a large $\text{x}_{\mathrm{zpf}}$ does not necessarily imply a large coupling rate. For instance, decreasing $L_{\text{x}}$ reduces $\text{x}_{\mathrm{zpf}}$ but increases $g_{\text{me}}$, while decreasing $L_{\text{y}}$ enhances both $\text{x}_\mathrm{zpf}$ and $g_{\text{me}}$. This behavior directly results from the strain factor $h/L_{\text{x}}^{2}$. Reducing $L_{\text{x}}$ decreases $\text{x}_{\mathrm{zpf}}$ but increases strain, which dominates and leads to larger coupling. This nontrivial scaling results from the first-order nature of the strain-mediated coupling in the membrane geometry, where the dominant contribution arises from the linear modulation of the in-plane strain rather than from higher-order geometric effects. Overall, suspended van der Waals (vdW) membranes benefit from both reduced device size and enhanced strain, with the dominant enhancement arising from the strain-induced first-order modulation of the in-plane lattice, which yields significantly stronger magnon–phonon coupling.

Figure 4 shows $g_{\text{me}}$ as a function of the initial pre-strain $h$ and the total membrane extension $L/L_{\text{x}}$, where $L$ is the deformed membrane length (see Eq.~B2). The coupling increases linearly with initial deformation, demonstrating that strain provides a simple and effective tuning knob. An initial deformation of $h \approx 45$ nm (corresponding to a total extension of $\sim 0.1\%$) enhances the coupling by approximately 1 kHz. Such deformations can be controlled in situ using a gate voltage~\cite{wang2010deflection}.

The performance of a magnomechanical platform is commonly characterized by the cooperativity, $C = \frac{4g_{\mathrm{eff}}^{2}}{\kappa_{\text{m}}\,\kappa_{\text{b}}}$,
where $\kappa_{\text{b}}$ and $\kappa_{\text{m}}$ are the mechanical and magnon linewidths, respectively.  
For the single-magnon regime ($\langle \hat{m}^{\dagger}\hat{m} \rangle = 1$), the effective coupling reduces to the single-magnon–phonon rate $g_{\mathrm{eff}} = g_{\text{0}}$, and the corresponding single-magnon cooperativity is $ C_{\mathrm{sm}} = \frac{4g_{\text{0}}^{2}}{\kappa_{\text{m}}\,\kappa_{\text{b}}}.$ Mechanical resonance frequencies of vdW membranes typically lie in the tens-of-MHz range, with phonon linewidths of a few kHz~\cite{makars}. Magnon linewidths on the order of hundreds of MHz have been reported for \ch{CrGeTe3} \cite{cgtms}.  Table~I compares the magnomechanical parameters and resulting cooperativities of our suspended \ch{CrGeTe3} membrane with those reported for a YIG sphere~\cite{cavitymagno,clinton} and a suspended \ch{CoFeB} beam~\cite{heikkila}. For the parameters shown in Table 1, $C_{\text{sm}}$ in \ch{CrGeTe3} membrane can be many orders of magnitude higher than that in YIG spheres.

To assess the feasibility of the proposed architecture, finite-element simulations were performed using COMSOL Multiphysics. The simulated system consists of a ferromagnetic membrane with voltage-tunable pre-strain. These simulations provide a realistic estimate of the voltage dependence of both the mechanical eigenfrequency and the strain overlap integral. A detailed description of the model and simulation results is provided in Appendix C; here, only the main results are summarized. For a 20 nm $\times$ 2 $\mu$m $\times$ 2 $\mu$m membrane suspended 285 nm above an electrode and subject to an initial in-plane tension of 0.3 N/m, the application of 20 V results in an out-of-plane displacement $h$ of approximately 20 nm. Simultaneously, the electrostatic force modifies the mechanical restoring potential, leading to an approximately twofold increase in the resonance frequency at 20 V. Under these conditions, the resulting single magnon-phonon coupling reaches approximately 0.1 kHz for a deflection of $20$ nm. This value is about four times smaller than that predicted by the analytical model, and around three orders of magnitude larger than the coupling rate observed in CoFeB beams and YIG spheres.

The two approaches agree qualitatively, i.e., the coupling vanishes in the absence of deformation and increases monotonically with $h$. We emphasize, however, that the two models describe different physical scenarios. The analytical model treats an already deformed membrane of deformation $h$, whereas the COMSOL simulations model the experimentally relevant case in which $h$ is generated by an electrostatic gate voltage. In the latter scenario, the gate simultaneously induces the deformation and modifies the mechanical eigenfrequency, effects that are absent in the analytical model. Further, the non-uniform pressure from the electrostatic gating can lead to deviations from the fundamental mode in the simulations. Together, these differences account for the approximate factor-of-four discrepancy at $h = 20$~nm, which becomes more 
pronounced at higher voltages as the frequency shift grows.

\section{Conclusions}
In this work, we have investigated the performance of suspended 2D van der Waals magnets as a platform to realize large magnon-phonon coupling. We examine the single magnon–phonon coupling rate in suspended membranes of two-dimensional vdW ferromagnets, deriving general expressions and presenting quantitative estimates for \ch{CrGeTe3}. By modeling a suspended membrane with clamped ends and in-plane magnetization, we confirm that a non-zero magnomechanical coupling requires an initial pre-strain deformation perpendicular to the membrane, and a non-zero component of magnon oscillation along the membrane's length. Our analysis shows that the primary mechanism in two-dimensional van der Waals magnets is the magnetoelastic interaction. We derive an analytical expression for the single magnon-phonon coupling rate and show that it increases linearly with the initial deformation height. The out-of-plane deformation generates a first-order in-plane tensile strain that couples directly to the magnetic excitations, leading to an enhanced effective interaction. The reduced dimensions increase the zero-point fluctuations, further increasing the coupling. This underscores the central role of bending-induced pre-strain in enabling and continuously tuning the magnomechanical coupling strength. For experimentally relevant parameters, the coupling rate can reach values up to a few kHz, which is more than three orders of magnitude higher than those observed in previously investigated platforms like YIG spheres and \ch{CoFeB} beams.

These findings establish suspended 2D magnetic membranes as highly promising platforms for large and tunable magnomechanical interactions. While the existing literature on hybrid quantum technologies has primarily focused on optomechanical systems \cite{quantumem, quantumtel}, the underlying concepts are directly transferable \cite{zuo2024cavity, magnomechquantum}. This opens a route toward magnon-based analogs of optomechanical devices, with potential applications in quantum memory \cite{magnomechquantum}, transduction \cite{hatanaka2022chip, optimaltrans}, and information processing. Our results, therefore, provide design principles for integrating 2D magnets into hybrid quantum systems and suggest new opportunities for magnonics in the quantum domain.

\section{Acknowledgements}
This project is a part of the projects "Ronde Open Competitie ENW pakket 21-3" (file number OCENW.M.21.215) and “Ronde Open Competitie XL” (file number OCENW.XL21.XL21.058), which are financed by the Dutch Research Council (NWO).

\appendix

\section{Magnons (Kittel mode)}
The Heisenberg Hamiltonian of a ferromagnet in the presence of an external magnetic field is given by
\begin{equation} \label{1}
    H = -\sum_{j,\sigma} \mathcal{J}_{j,j+\sigma}\Vec{S}_j \cdot \Vec{S}_{j+\sigma}
   -\sum_{n,j} \mathcal{D}_n(S^n_j)^2
  -\sum_j{ \gamma \hbar} \Vec{S_{j}}\cdot\Vec{B},
\end{equation}
where $\Vec{S}_{j}$ is the spin (in units of $\hbar$) at lattice site $j$, $S^{n}_{j}$ the component along the $n$-axis ($n \in {\text{x},\text{y},\text{z}}$), $ \mathcal{J}_{j, j + \sigma}$ is the exchange interaction between spins at lattice points $j$ and $j+\sigma$, $\mathcal{D}_{n}$ is the strength of the magnetocrystalline anisotropy. $\vec{B}$ is an external magnetic field that can be applied in any general direction that is in the plane, and $\gamma$ is the gyromagnetic ratio given by, $\gamma/2 \pi$ = $28$ GHz/T. In the uniform Kittel mode, the spins precess in-phase allowing us to define the macrospin operator
\begin{equation}
\vec{S} = \sum \vec{S_{j}} = S^{\text{x}} \hat{\text{x}} + S^{\text{y}} \hat{\text{y}} + S^{\text{z}} \hat{\text{z}} = V_{\text{m}} \vec{M}/\gamma,
\end{equation} 
where $\vec{M}$ is the total magnetization, and $V_{\text{m}}$ is the volume of the magnetic material. Assuming that the external magnetic field is applied along $\text{y}$ and completely aligns the spins in-plane along $\text{y}$, the spins can be written in terms of the magnon creation and annihilation operators $(\hat{m}^{\dagger}, \hat{m})$ by the Holstein-Primakoff (HP) transformation as \cite{rezende2019introduction,stancilprabhakar},
\begin{equation}
    \begin{aligned}
        &S^{+} = \sqrt{S}\left(1 - \dfrac{\hat{m}^{\dagger}\hat{m}}{2s} \right)^{1/2} \hat{m}, \\ 
        &S^{-} = \sqrt{S}\left(1 - \dfrac{\hat{m}^{\dagger}\hat{m}}{2s} \right)^{1/2} \hat{m}^{\dagger},\\
        &S_\text{y} = S - \hat{m}^{\dagger}\hat{m},
    \end{aligned}
\end{equation}
where $S^{+} = S^\text{x} + iS^\text{z}$, and $S^{-} = S^\text{x} - iS^\text{z}$ are the spin raising and lowering operators, respectively. Linearizing Eq. (A3) and substituting in Eq. (A2), we obtain Eq. (4) in the main text.

The Hamiltonian in Eq. (A1) can be diagonalized \cite{cgtfmrtheory, cgtms} to calculate the ferromagnetic resonance frequency, i.e., the frequency of the magnons in the Kittel mode.

\begin{figure*}
 \begin{center}
    \begin{subfigure}{0.53\textwidth}
    \begin{overpic}[width=\linewidth]{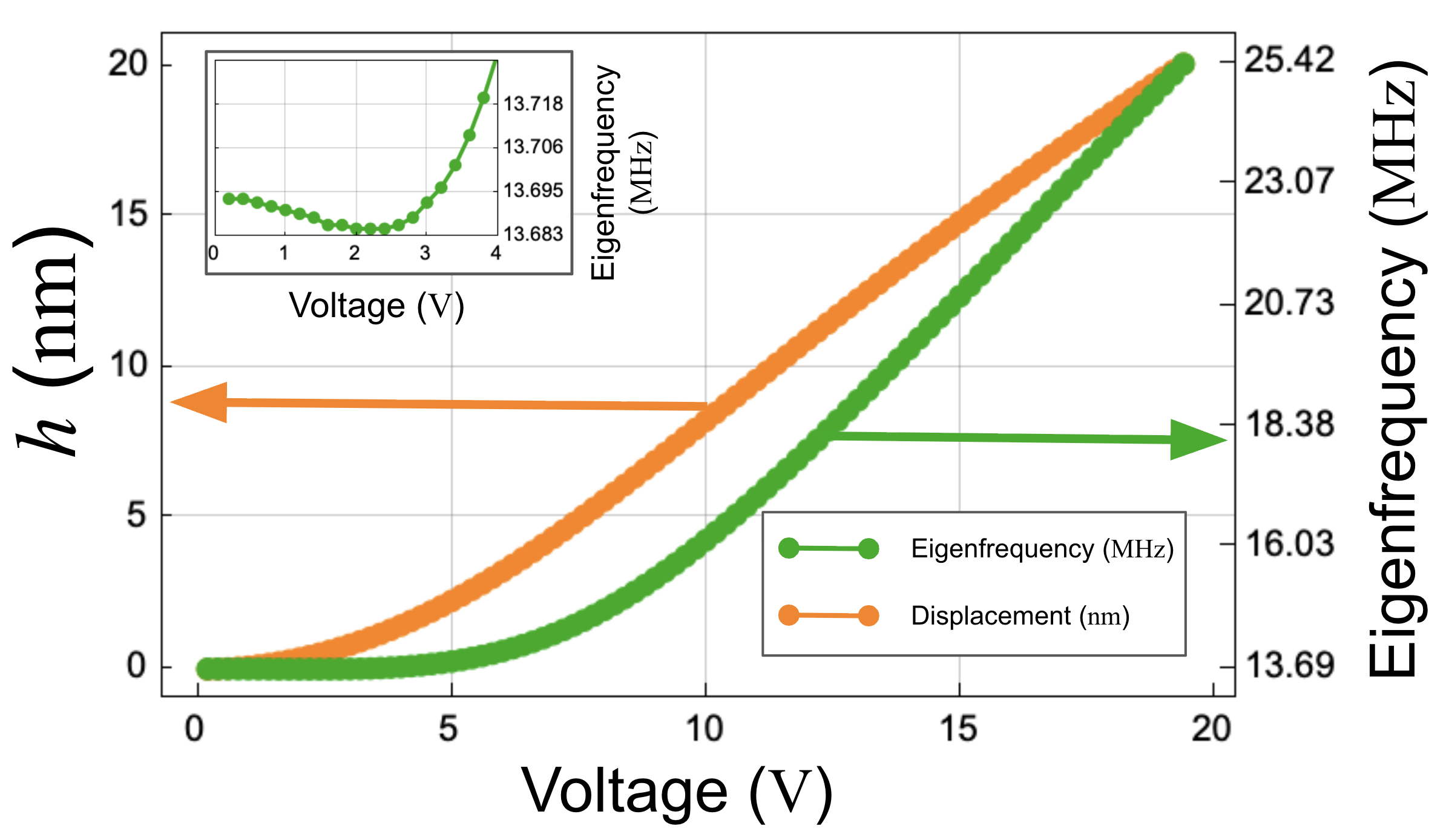}
        \put(0,54){\large (a)}
    \end{overpic}
  \end{subfigure}
  \begin{subfigure}{0.46\textwidth}
    \begin{overpic}[width=\linewidth]{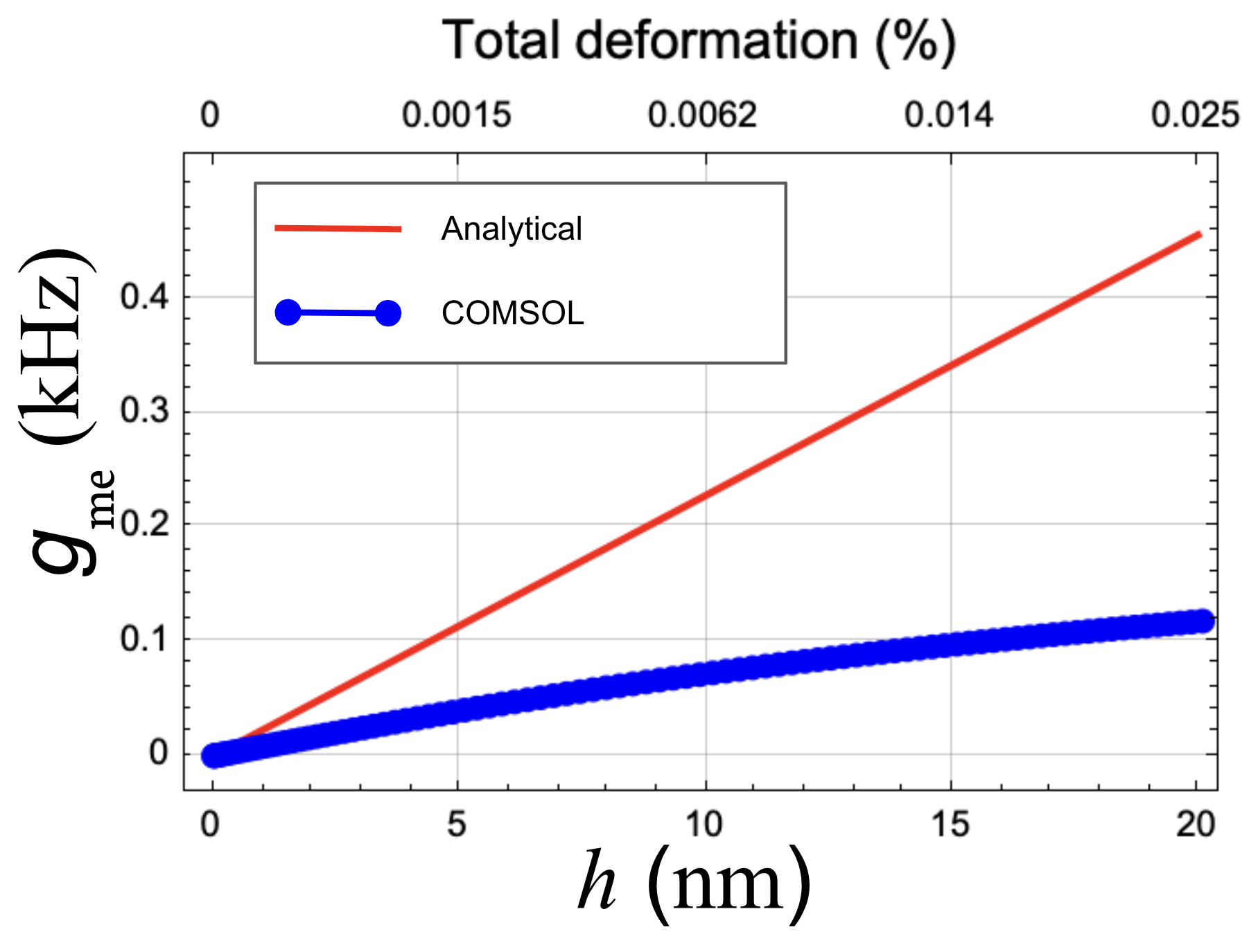}
        \put(4,62){\large (b)}
    \end{overpic}
  \end{subfigure}
  \end{center}
 \caption{(a) Mechanical deformation $h$ and mechanical eigenfrequency as a function of applied gate voltage, obtained from COMSOL simulations. The inset shows eigenfrequency as a function of voltage from 0 to 4 V (b) Single magnon-phonon coupling strength as a function of the mechanical deformation $h$. The red line shows the analytical result, which neglects the reduction of the mechanical eigenfrequency induced by electrostatic gating. The blue symbols correspond to COMSOL simulations.}
\end{figure*}

\section{Strain}
As discussed in section II, the system is pre-strained such that there is a deformation of height $h$. The suspended ferromagnet is displaced perpendicular to its length and clamped at both ends, $(-L_{\text{x}}/2,0)$ and $(L_{\text{x}}/2,0)$. In the lowest mechanical mode, the resulting out-of-plane displacement is given by (see Figure 1)

\begin{equation}
\text{r}_{\text{z}}\text{(x},t) = \text{r}_{\text{z}}(t) \cos\left(\frac{\text{x} \pi}{L_\text{x}}\right), \quad \text{where } \text{r}_{\text{z}}(t) = h + \Delta \text{z} e^{i \omega t}.
\end{equation}
The local deformation of the system along x can be approximated for small displacements ($h, \Delta \text{z} \ll L_\text{x}$) by  
\begin{equation}
d\text{r}_{\text{x}} = d\text{x} \left( \sqrt{1 + \text{r}_{\text{z}}^{2}(t)\dfrac{\pi^2}{L_{\text{x}}^2}\sin^{2}\left[\dfrac{\pi \text{x}}{L_{\text{x}}} \right]} -1 \right).
\end{equation}
The total strain along x for small displacements without normalization can be approximated as
\begin{equation}
\epsilon_{\text{xx}}(t) \approx \frac{\pi^2}{2 L_\text{x}^2} \sin^{2}\left[\dfrac{\pi \text{x}}{L_{\text{x}}} \right]\left(h + \Delta \text{z} e^{i \omega t}\right)^2.
\end{equation}
Expanding and only retaining the terms linear in $\Delta \text{z}$, the strain simplifies to  
\begin{equation}
\epsilon_{\text{xx}}(t) \approx \frac{\pi^2}{2 L_\text{x}^2} \sin^{2}\left[\dfrac{\pi \text{x}}{L_{\text{x}}} \right] \left(2 h \Delta \text{z} e^{i \omega t}\right).
\end{equation}
Finally, quantizing the strain by replacing the displacement amplitude $\Delta \text{z}$ with $\text{x}_{\text{zpf}}$ and displacement operators \cite{clinton}, we obtain  
\begin{equation}
\epsilon_{\text{xx}}(t) = \frac{\pi^{2} h}{L_\text{x}^2} \sqrt{\dfrac{\hbar}{2m_{\text{eff}} \omega_{b}}}\sin^{2} \left[ \dfrac{\pi \text{x}}{L_{\text{x}}} \right] (\hat{b}^\dagger + \hat{b}) = \epsilon^{\text{max}}_{\text{xx}} (\hat{b}^{\dagger} + \hat{b}).
\end{equation}
Eq. (B5) is Eq. (5) in the main text.

\section{Electrostatic tuning of magnon-phonon coupling}

To assess the experimental feasibility of tuning the magnon–phonon coupling via an electrostatic gate, we perform finite-element simulations using COMSOL Multiphysics. The simulations model a suspended membrane of dimensions $2 \times 2$ $\mu \text{m}^2$ and thickness $20$ nm, subjected to an electrostatic force arising from a voltage applied between the membrane and an electrode located a distance $d$ (depth of the vacuum gap, which we set to 285 nm) below it, forming a vacuum-gap capacitor. When opposite charges accumulate on the membrane and the electrode, the resulting electrostatic attraction pulls the membrane downward, inducing a static displacement.

The electrostatic pressure exerted on the membrane is given by \cite{Senturia2005,Lardies2010}
\begin{equation}
P_e = \frac{\epsilon_0 V^2}{2\bigl(d - \text{r}_\text{z}\bigr)^2},
\end{equation}
where $\epsilon_0$ is the vacuum permittivity and $V$ is the applied voltage.

The COMSOL model is constructed using the membrane physics interface, assuming a linear elastic material with geometric nonlinearity enabled. We include a uniaxial pretension perpendicular to the supports. The electrostatic pressure is computed iteratively by COMSOL and applied as a face load on the undeformed surface. As the membrane deflects, the pressure is updated self-consistently according to Eq.~(C1). To determine the voltage-induced static strain, we perform a parametric stationary sweep. The mechanical eigenmodes are then obtained from an eigenfrequency study that uses the stationary solution as its operating point.

The electrostatic force pulls the membrane toward the gate, producing a static out-of-plane deformation $h$ that breaks the symmetry of the strain field. Figure 5 (a) shows the resulting membrane deformation and the fundamental mechanical eigenfrequency as functions of the applied gate voltage. The deformation increases monotonically with voltage, demonstrating that the membrane displacement can be continuously and reversibly controlled by electrostatic gating.

In addition to inducing static deformation, the electrostatic force modifies the mechanical restoring potential. At first, as the gate voltage increases, the effective spring constant of the membrane is reduced due to electrostatic spring softening \cite{Steeneken_2021}, leading to a decrease in the mechanical eigenfrequency, as shown in Figure 5 (a) from 0 to approximately 2 V (see Figure 5 (a) inset). This behavior is a well-known feature of electrostatically actuated nanomechanical resonators. Then, from 2 to 20 V, the static in-plane strain that is generated by the vertical displacement produces a tension that overcomes the effect of electrostatic softening and increases the frequency in an approximately linear fashion \cite{voltagetuning}. The mode shape obtained from COMSOL as a function of the applied voltage deviates from the idealized profile of the fundamental assumed in the analytical model. In particular, the presence of nonuniform electrostatic pressure, as well as bending effects, leads to distortions of the displacement profile. These deviations result in reduced in-plane strain observed in the COMSOL simulations compared to the analytical prediction, further reducing the coupling.

Figure 5 (b) compares the single magnon–phonon coupling strength obtained from COMSOL simulations with the analytical prediction derived in Eq.~10. The analytical model assumes a constant mechanical eigenfrequency and therefore neglects electrostatic force effects, whereas the numerical simulations fully account for the voltage-induced change of the eigenfrequency. As a result, the analytical theory overestimates the coupling strength.

Despite this quantitative discrepancy, both approaches show the same qualitative behavior: the single magnon–phonon coupling vanishes in the absence of deformation and increases monotonically with increasing $h$. The agreement in scaling confirms that electrostatic gating provides a practical and reliable method for tuning the magnomechanical coupling strength in suspended van der Waals ferromagnets.
 
\bibliography{apssamp}
\end{document}